\documentclass[aps,preprint,superscriptaddress,showpacs]{revtex4}

\usepackage{graphicx}
\usepackage{amsmath}
\usepackage{amssymb}
\usepackage{hyperref}
\bibstyle{apsrev.bib}

\begin{document}

\title{Fick's law and Fokker-Planck Equation in inhomogeneous environments}

\author{F. Sattin}\email{fabio.sattin@igi.cnr.it}

\affiliation{Consorzio RFX, Associazione EURATOM-ENEA sulla fusione, Corso Stati Uniti 4,
Padova, Italy}

\begin{abstract}

In inhomogeneous environments, the correct expression of the diffusive flux
 is not always given by the Fick's law $\Gamma = - D \nabla n $.
The most general hydrodynamic equation modelling diffusion is indeed the 
Fokker-Planck Equation (FPE). The microscopic dynamics of each specific system
may affect the form of the FPE, either establishing connections between the 
diffusion and the convection term, as well as providing supplementary terms.   
In particular, the Fick's form for the Diffusion Equation may arise only in consequence 
of a specific kind of microscopic dynamics. It is also shown how, in the presence of sharp inhomogeneities,
even the hydrodynamic FPE limit may becomes inaccurate and mask some features of the true solution, 
as computed from the Master Equation.

\end{abstract}


\maketitle

\textit{Introduction.}
The fluid modelling of the time- and space evolution
of quantities within complex environments, whose dynamics may only be
treated on statistical grounds, is made using the diffusion equation (DE)
$\partial_t n = D \partial_x^2 n$. 
This phenomenological equation arises from two more fundamental equations: the continuity
equation for $n$: $ \partial_t n  = - \partial_x \Gamma$, 
and the Fick's law (or Fourier's law) \cite{ref1}
$\Gamma = - D\partial_x n$,
where $x$ and $t$ are the spatial
coordinate and the time, respectively; the diffusivity $D$ is
a constant dependent from the medium. A pedagogical overview of
Fick's (Fourier's) law and diffusion equation may be found in \cite{ghosh}.\\
The postulate of homogeneity may
hold just as a first-order approximation, whereas most systems must ultimately
allow for some degree of non-uniformity. Almost unavoidably, therefore, one is
faced with the question: how DE
has to be generalized to such systems. The exact answer to this question is of
relevance for a plethora of problems in practically any branch of natural
sciences: from physics, to chemistry, geology, biology, social sciences, \ldots{}. \\
Heuristically, the difficulty related to the generalization of DE
may be understood as follows: an inhomogeneous environment should make
$D$ position-dependent: $D \to D(x)$. 
There are, however, several choices for $\Gamma{}$ that differ when
$D = D(x)$, but that collapse to the same identical form 
when $D$ is constant. Therefore, the problem may be restated as: what is the correct
generalization of Fick's law (provided that one exists) in inhomogeneous environments.\\
This subject appears repeatedly addressed in literature; however, it is
difficult to find the explicit exposition of a general solution.   
In Van Kampen's book \cite{ref2}, 
it is argued that one cannot decide \textit{a priori} what the correct form
for $\Gamma{}$ is, which rather depends upon the properties of
the problem studied. 
Landsberg (\cite{ref3} and references therein), points out that, to some extent, it 
is a matter of convention, provided that supplementary (convective) terms are
added suitably. In other terms, the definition of a diffusive and a convective
flux is not univocal, only the total flux is. The paper providing the clearest
intuitive insight and at the same time detailed calculations about what goes on
in such situations is probably Schnitzer's \cite{ref4}.
We mention also the papers \cite{ref5,ref5bis}, featuring computer experiments 
 and presenting further bibliography about this subject. 
Papers \cite{ref5ter,ref6,ref7}
feature analytical and experimental work, demonstrating that the straightforward 
generalization of Fick's law $ \Gamma = - D(x) \partial_x n(x)$  cannot hold in all systems. 

In order to quantitatively address the issue, it is necessary to deal with a 
reasonably accurate modelling of the dynamics at the microscopic level: transport equations,
thus, will emerge at the level of large length scales.
The tool we adopt is provided by the Master Equation (ME):
\begin{equation}
\frac{{\partial n(x,t)}}{{\partial t}} =  - \frac{{n(x,t)}}{{\tau (x)}} + \int
{dx'p(x - x',x')\frac{{n(x',t)}}{{\tau (x')}}}
\label{eq5}
\end{equation}
ME (\ref{eq5}) yields a coarse grained probabilistic description of a microscopic system
driven by a Markov process, and can be visualized as the continuity equation for the passive scalar 
quantity \textit{n(x,t)} (which, properly speaking, is a probability density) subject to transitions (\textquotedblleft{}jumps\textquotedblright{}) modifying
its state from $x'$ to $x$, with probability
$p(x - x',x')$, and at a rate
$1/\tau{}(x)$ (see chapter 1 of \cite{langevin}). 
Equation (\ref{eq5}) contains virtually all the solutions of the transport problem, once the
functions $p$ and $\tau{}$ are given. On the other hand, it is often unpractical to deal
directly with it, particularly in higher-dimensional problems. Therefore, and particularly if
a clear-cut separation of scales exists in the problem studied, it is customary to take its
long-wavelength limit, which washes out details at the finest scales and turns the integral
equation (\ref{eq5}) into a famous differential equation: the Fokker-Planck Equation (FPE)
(see, e.g., chapter 9 of \cite{balescu}):
\begin{equation}
\frac{{\partial n(x,t)}}{{\partial t}} =  - \frac{\partial }{{\partial x}}\left(
{U(x) n} \right) + \frac{{\partial ^2 }}{{\partial x^2 }}\left(
{D(x)n} \right)
\label{FPE}
\end{equation}
Within the ME formulation, all the physics is built into the functions $p$ and $\tau$. 
In the passage from ME to FPE, $p$ and $\tau$ are packed into the diffusive and convective terms, $D, U$. 
Therefore, the analytical expression of $D,U$, ultimately relies on the constraints that 
the problem to be solved places on $p, \tau$. Is it possible, basing upon general considerations
on the microscopic dynamics, to identify equivalent classes of systems, that is, 
systems that lead to the same qualitative form of the FPE ?
As we shall show later, the initial question advanced in this Introduction is related to
this point: the Fick's form of the diffusion equation is a particular limiting case of the FPE, 
that arises when the microscopic dynamics fulfils a given simmetry. \\ 
The purpose of this paper is to provide a discussion about this topic. Furthermore, 
we will address the broader issue of the validity of the scale separation at the 
basis of the FPE. We will show that, whenever, this hypothesis is not fulfilled,
additional terms to the FPE need to be considered.   \\
\textit{From the Master Equation to the Fokker-Planck Equation. } 
The simplest way to pass from ME to FPE is by expressing the integrand in
Eq. (\ref{eq5}) in terms of the small parameter $\Delta{} =
x' - x$, which is of order the mean jumping length $L_p$: 
\begin{equation}
{p(x-x',x') \over \tau(x')} n(x') =  {p(-\Delta,x+\Delta) \over \tau(x+\Delta)} n(x+\Delta)
\end{equation}
and expanding around $x$ in powers of $\Delta$ (Kramer-Moyal expansion). 
However, this step is justified provided that $p,\tau, n$, are not strongly varying functions
of $x$ over distances of order $L_p$. If we assume that $n$ is a smooth function of $x$,
we may concentrate on the other quantity: $h = p/\tau$.
A branching into two cases is possible:
\textit{(1)} $h$ is a smooth function, or
\textit{(2)} $h$ is not. Although, condition 
\textit{(2)} actually contains \textit{(1)} as a particular case, it turns out
convenient to consider them separately, since \textit{(1)} is easier
to deal with. \\
Finally, we will consider also the case \textit{(3)}, when $n$ itself is not a smooth function.  \\
\textit{Case \textit{(1)}: both $n$ and $h$ are smooth functions.} 
We are allowed to make
a Taylor expansion in powers of the function $h \times n$. 
The result, truncated to second order, yields Eq. (\ref{FPE})
with
\begin{equation}
U  = \int {d\Delta \frac{{p(\Delta ,x)}}{{\tau (x)}}} \Delta \,,\quad 
D = {1 \over 2} \int {d\Delta \frac{{p(\Delta ,x)}}{{\tau (x)}}} \Delta ^2
\label{eq7}
\end{equation}
Limiting the truncation to second order is ordinarily 
justified on the basis of Pawula theorem \cite{ref8,ref9}.\\ 
All the information relevant to our problem is packed into
$U, D$.  Two important cases are  
(\textit{A}) $U = (dD/dx)$, or
(\textit{B}) $U = 0$. Case (\textit{A}) recovers Fick's law, while case 
(\textit{B}) yields the solution 
\begin{equation}
\partial_t n = \partial^2_x (D(x) n(x))
\label{eq4}
\end{equation}
Both results may be verified by direct substitution into Eq. (\ref{FPE}).
It turns out that relation (\textit{A}) arises straightforwardly from ME (\ref{eq5})
by postulating the simmetry
\begin{equation}
\frac{{p(x' - x,x)}}{{\tau (x)}} = \frac{{p(x - x',x')}}{{\tau (x')}} \to
\frac{{p(\Delta ,x)}}{{\tau (x)}} = \frac{{p( - \Delta ,x + \Delta )}}{{\tau (x +
\Delta )}}
\label{eq8}
\end{equation}
which ensures the time reversal symmetry of the microscopic dynamics. Indeed, a
first-order Taylor expansion of the second argument around
$x$ yields, after rearranging,
\begin{equation}
\Delta \frac{{dp( - \Delta ,x)}}{{dx}} = p(\Delta ,x) - p( - \Delta ,x) + \Delta
{\kern 1pt} p( - \Delta ,x)\frac{{d\ln \tau }}{{dx}}
\label{eq9}
\end{equation}
Using Eq. (\ref{eq9}) into the integrals (\ref{eq7})
yields the sought result \textit{(A)} (For a different derivation, see Prof. Feder's lecture notes \cite{federURL}). \\
The solution (\textit{B}) has some relevance, too, since it corresponds to the choice of a symmetrical
kernel: $p(\Delta,x) = p(-\Delta,x)$. Although apparently natural, the range of validity
of this condition is actually rather narrow, as it cannot hold under smoothly varying conditions, 
where $ p(\Delta,x) \neq p(-\Delta,x)$; that is, the probability
for a particle of jumping rightwards or leftwards cannot be the same.
In order to better understand this point, let us consider a system where test particles 
collide against some scattering centres.
Jumps are arcs of ballistic motion between two collisions. If the system is not
homogeneous the density $n_{sc}$ of the scattering centres is not uniform. Let us suppose,
say, $dn_{sc}/dx < 0$: a test particle at $x$ has a 
larger probability of striking a scattering centre that is on its left ($x -\delta x$)
rather than on its right ($x + \delta x$), and therefore of being backscattered in the opposite direction. 
Hence, there is a larger probability
of bouncing back rightwards than the converse. \\
Having ruled out the case \textit{(B)} for several inhomogeneous systems, one could wonder how general
is condition \textit{(A)}. It turns out that \textit{(A)}
generically holds for a large class of 1-degree-of-freedom Hamiltonian systems \cite{ref10,ref11,ref12}
(see also \cite{feder1} for another particular case).
For more general systems, and especially in systems with more degrees of freedom,
the above constraints between $U$ and $D$ cannot be guaranteed to hold
any longer, and FPE may allow in principle for a wide variety of cases. Two 
such instances, recalled in \cite{ref10}, are: the self-consistent motion of charged particles
in a set of Langmuir waves, and the 2-dimensional guiding-centre motion of a particle in a 
varying stochastic electrostatic field. \\
\textit{Case \textit{(2)}: $n$ is a smooth function but $h$ is not. } 
Let us consider case \textit{(2)}, when $p$ and/or $\tau$ present sharp
variations: we mean they vary on scales smaller than
$L_{p}$. This is the situation when one needs modelling
systems characterized by sudden transitions between regions with widely different
physical properties. Therefore one is forced to study the case when 
we can still expand $n$ in powers, but now must leave $h$
unexpanded: after some calculations
\begin{equation}
{\partial n \over \partial t} = {\partial ^2 \over \partial x^2 } \left(\hat{D} n\right) - 
{\partial \over \partial x} \left(\hat{U} n \right) + \hat{W} n
\label{eq20}
\end{equation}
\begin{eqnarray}
\hat{D} &=& { \hat{m}_2 \over 2} \nonumber \\ 
\hat{U} &=& \hat{m}_1 + \partial_x \hat{m}_2 \nonumber \\
\hat{W} &=& {\partial_x^2 \hat{m}_2 \over 2} + \partial_x \hat{m}_1 + \hat{m}_0 \nonumber \\
\hat m_j  &=& \int {dx'(x - x')^j \frac{{p(x - x',x')}}{{\tau (x')}}}  - 
\delta_{j0} \frac{1}{{\tau(x)}} \;\; (j = 0, 1, 2) \nonumber
\end{eqnarray}
We have now an ``extended'' Fokker-Planck Equation, due to 
the presence of an additional $\hat{W} n$ term. One may feel unconfortable about the presence
of this term, since at first sight it appears to spoil the conservation of matter: 
$ \partial N /\partial t  =  \partial \int n /\partial t = 0$, which is built into Eq. (\ref{eq5}) and 
Eq. (\ref{FPE}). By integrating Eq. (\ref{eq20})
over $x$, instead, one has  $ \partial N/\partial t = \int \hat{W} n(x) dx $, which is not granted
a priori to be zero. We will show, instead, that it is exactly the case: the role of $\hat{W}$ is that 
of transferring matter from one point to another, 
rather than that of a net sink or source.  
Heuristically, it may be guessed on the basis of the fact that Eq. (\ref{eq20}) is just 
an intermediate passage in the chain of calculations leading from Eq. (\ref{eq5}) to Eq. (\ref{FPE}). 
Since both the starting and the end expressions do conserve matter, also the intermediate step must. 
For simplicity, from here on, we will consider the case of constant $\tau = 1$; hence, 
only variations in $p$ will be dealt with. Let $x_0$ be a point around which $p$ shows a sharp
variation. For the sake of simplicity, let us consider a step-like variation around $x_0 = 0$:
\begin{equation}
p(x-x',x') = \left\{ \begin{array}{*{16}c} {p_L(x-x') \quad x' < 0} \\
                                           {p_R(x-x') \quad x' > 0} \\
                     \end{array} \right.
\end{equation}
Since, far from $x_0$, the system is almost homogeneous, we may suppose that the jumping lengths
are symmetrical: $p_{L,R}(x-x') = p_{L,R}(|x-x'|)$.
We show now that, under these hypotheses, $\hat{W}$ reverses sign around $x_0 = 0$: $\hat{W}(-x) = - \hat{W}(x)$.
For brevity, we proceed to demonstrate only that $\hat{m}_0$ is an odd function of $x$. Calculations for the other 
two terms carry on along the same lines:
\begin{eqnarray}
\hat{m}_0(x) &=& \int dx' p(x-x',x') - 1 = \int_{x' < 0} dx' p_L(x-x') + \int_{x' > 0} dx' p_R(x-x') - 1  \nonumber \\
 &=& \int dx' p_L(x-x')+ \int dx' p_R(x-x') - \int_{x' > 0} dx' p_L(x-x')  - \int_{x' < 0} dx' p_R(x-x')- 1 \nonumber \\
 &=& 1 + 1 -  \int_{x' > 0} dx' p_L(x-x')  - \int_{x' < 0} dx' p_R(x-x')- 1  \nonumber \\ 
 &=& - \left(\int_{x' > 0} dx' p_L(x-x')  + \int_{x' < 0} dx' p_R(x-x')- 1 \right)
\end{eqnarray}
If in the last line we make the change of variables $ x \to - x , x' \to - x'$, we get that the 
term within parentheses is just $\hat{m}_0(-x)$. The second fundamental property of $\hat{W}$ is that $\hat{W}$ is different from zero only over a region around $x = 0$ of width a few  jumping lengths: It is straightforward to show that $\hat{m}_0 = d\hat{m}_1/dx = d^2\hat{m}_2/dx^2 = 0$ when $|x| >> L_p$ . 
On the other hand, $n$ is just a weakly varying function over distances of order the jumping length: this 
is a postulate implicit in the Taylor expansion done when going from Eq. (\ref{eq5}) to Eq. (\ref{eq20}). 
Therefore, combining the above results, $ \int \hat{W}(x) n(x) \; dx \approx n(0) \int \hat{W}(x) \; dx$, and $\int \hat{W} \; dx = 0$. This concludes the demonstration. \\
In order to be more quantitative, let us consider the paradigmatic case  
of a gaussian diffusion with sharp variations of the jumping length:
\begin{eqnarray}
p(x - x',x') &=& \frac{1}{{\sqrt {2\pi \sigma ^2 (x')} }}\exp \left[ { - \frac{{(x
- x')^2 }}{{2\sigma ^2 (x')}}} \right] \\
 \sigma (x) &=& \left\{
{\begin{array}{*{16}c}
{\sigma _L \;\;, \quad x < 0}  \\
{\sigma _R \;\;, \quad x > 0}  \\
\end{array}} \right. \nonumber
\label{eq21}
\end{eqnarray}
It is straightforward to compute $\hat{W}$ for this system:
\begin{eqnarray}
\hat{W} &=& \sqrt{\frac{1}{8 \pi }} x \left(\frac{e^{-\frac{x^2}{2 \sigma_R^2}} \left(4
   \sigma_R^2-x^2\right)}{\sigma_R^3}+\frac{e^{-\frac{x^2}{2 \sigma_L^2}} \left(x^2-4
   \sigma_L^2\right)}{\sigma_L^3}\right)  \nonumber \\ 
   &-& \frac{1}{2} \left( \text{erf}\left(\frac{x}{\sqrt{2} \sigma_L}\right)-
   \text{erf}\left(\frac{x}{\sqrt{2} \sigma_R}\right)\right) 
\label{eq:w}
\end{eqnarray} 
\begin{figure}
\includegraphics[width=70mm]{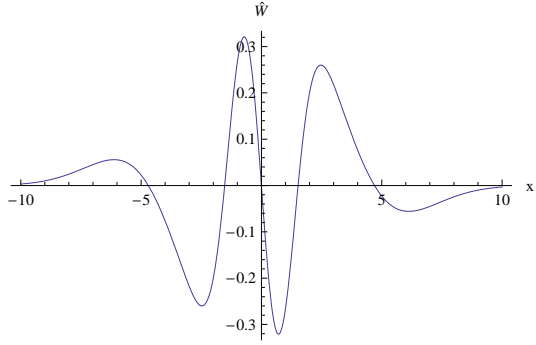}
\caption{$\hat{W}$ from Eq. (\ref{eq:w}), with $\sigma_L = 1$, $\sigma_R = 2.5$.}
\label{figw}
\end{figure}

\begin{figure}
\includegraphics[width=70mm]{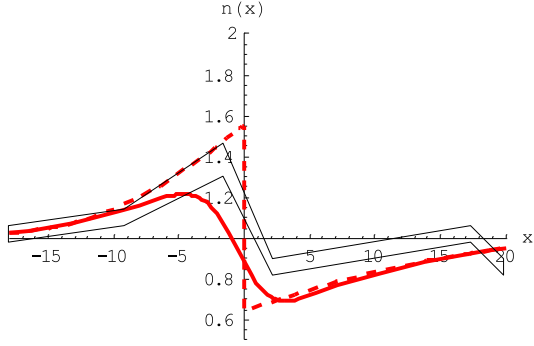}

\caption{(Color online) Solution of Eq. (\ref{eq20})
(solid line) and Eq. (\ref{eq4})
(dashed line) at $t = 32$. Starting profile is flat: $n(t = 0) = 1$.
The contoured region roughly envelops the area where experimental points
lie (adapted from Fig. 5 of Ref. \cite{ref7}). The two solutions are
computed for slightly different boundary conditions: Eq. (\ref{eq20})
has been numerically solved imposing $dn/dx = 0$ at $x = \pm 40$. 
Eq. (\ref{eq4}) may be integrated analytically.
Its solution is discontinuous in $x = 0$, but the flux
$d(D n)/dx$ is continuous \cite{ref7}. 
For this function, $ dn/dx \to 0$ only
asymptotically as
$\vert{}x\vert{} \to \infty{}$.}
\label{fig1}
\end{figure}

Its profile is given in figure (\ref{figw}). \\
We have available an excellent experimental test bench of this result: 
paper \cite{ref7} presents a study of tracer diffusion between gelatine
solutions with different viscosity, which means different effective diffusivity. The width of the interface between the two solution is very small, and can be considered as zero for our 
purposes. Hence, the whole system may be modelled as two
regions with different jumping lengths, just done above.
It is apparent at this point that applying Eq.
(\ref{FPE}) at this problem
is of dubious validity, since $p$ is discontinuous at
$x' = 0$ in its second argument. Nevertheless, just as an
exercise, we may formally evaluate $U, D$, for this
choice of $p$:
\begin{equation}
U  = 0\,,\quad D  = {1 \over 2} \left\{ {\begin{array}{*{20}c}
{\sigma _L^2 \; , \quad x < 0}  \\
{\sigma _R^2 \; , \quad x > 0}  \\
\end{array}} \right.
\label{eq22}
\end{equation}
thereby recovering Eq. (\ref{eq4}).
$ \hat{W} $ appearing in (\ref{eq20}) has been computed in Eq. (\ref{eq:w}), and 
$\hat{D}, \hat{U}$ are analytically computable alike, although 
we do not provide their complete expression here for saving space. 
The numerical values we choose for $\sigma_{L,R}$ are
$\sigma_{L} = \sqrt{2.4}$, $\sigma_{R} = \sqrt{5.8}$, 
and solve for the time evolution of $n(x,t)$ starting from a flat profile:
$n(x,t = 0) = 1$. In Fig. (\ref{fig1})
we show the spatial profile at a later time for the solution of Eq.
(\ref{eq20}) as well as of Eq. (\ref{eq4}).
These profiles should be compared against their counterpart from experiments
(Fig. 5 in ref. \cite{ref7}). It becomes apparent that the smooth transition around 
$x = 0$, that exists in real data, is completely masked by the use of
Eq. (\ref{eq4}), while is correctly recovered by Eq. (\ref{eq20}).\\

\begin{figure}
\includegraphics[width=70mm]{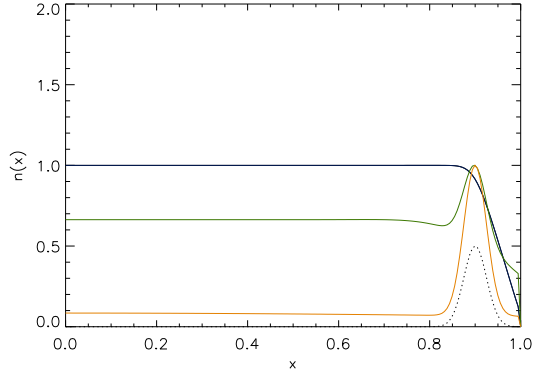}

\caption{(Color online) Density profiles $n$ due
to the source (dashed curve) for three values of $L_{p} =
0.0125$ (black curve), 0.125 (green curve), 1 (orange curve).  Here, the container
has size $L = 1$, and the source width is
$L_{s} = 0.025$. Reflecting conditions are imposed at $x = 0$ and absorbing boundaries
at $x = 1$.}
\label{fig2}
\end{figure}

\textit{Case \textit{(3)}: $n$ is not a smooth function and the full Master Equation is needed. }
This case acquires relevance
when variations in $n$ occur over scales smaller than the jumping length: $L_{p} > L_{n} \equiv{} n/(dn/dx)$.
In unbounded not driven systems this criterion may be satisfied only transiently,
starting from highly localized profiles. Left to itself, density relaxes so as to fulfil
$L_{n} > L_{p}$. However, $L_{n}$ may be bounded by imposing
absorbing boundaries. Hence, $L_{n} < L$, where $L$
is the system' size. Since absorbing boundaries imply loss of density from the
system, in order to maintain a steady state it is necessary to add a source, which is
parameterized by its spatial extension, and hence a further typical length, $L_{s}$. The usual
ordering is $L_{p} <  L_{s} < L $, while now we investigate the reversal of
this ordering. For simplicity, we will limit to consider the steady state
$\partial {\kern 1pt} /\partial t = 0$. Eq. (\ref{eq5})
with a source $S$ can be solved in Fourier space for
$g(x) = n(x)/\tau{}(x)$:
\begin{equation}
g(k) = \frac{{S(k L_s)}}{{1 - p(k L_p)}}
\label{eq23}
\end{equation}
In (\ref{eq23}) we have made it explicit that $p$ depends upon scale length $L_p$, and $S$ upon $L_s$.
For $k L_{p}, k L_{s} << 1$, one recovers the solution of the diffusion
equation if $p$ is a smooth symmetrical function: $1 - p(k) \approx k^2, \, k \to 0$.
(If $p$ lacks mirror symmetry, a convective term appears). 
Hence, over very large spatial scales one does not expect any novel
feature to arise. However, when $k L_{s} \approx 1$ but $k L_p >> 1$
the denominator of (\ref{eq23}) is almost one: $p$ must have finite support,
thus $p \to 0, k \to \infty$. Hence
\begin{equation}
g(k) \approx S(k)
\label{eq24}
\end{equation}
Close to the source, and over distances of order of the source's
width, the density profile matches exactly that of $S$. This is an effect
totally unpredictable within the FPE formulation, that instead would smear the
density throughout the whole system. In order to check numerically this
prediction, we have solved Eq. (\ref{eq5})
for a given source profile and different values of
$L_{p}$. The results, fully confirming our analytical
estimates, are shown in Fig. (\ref{fig2}).\\
\textit{Conclusions. }
In summary, inhomogeneity may have dramatic effect on the modelling of the spreading
of some quantity into a medium. It affects the writing of the equation of motion,
turning it into a problem that has unambiguous solutions, but generally not valid ones
for all classes of systems. Two fundamental criteria are identified. From the one hand, 
the microscopic dynamics imposes constraints between the diffusive and convective 
coefficients of the FPE. On the other hand, the form of the FPE itself depends on the
existence of scale separation between typical lengths existing in the system: when 
this criterion is fulfilled and transport scales are clearly smaller than any other scale,
the classical FPE is valid. On the opposite side, when no clear separation can be made, 
additional terms need to be added into an ``augmented'' FPE, till to the extreme case, where
only the full ME may provide correct results. The first criterion is obviously the most important: 
FPE is often used for modelling purposes starting from a limited 
knowledge of the underlying microscopic dynamics. Clearly,
one cannot establish {\it a priori} if scale separation does hold in the problem at hand.
In these situations, therefore, the standard FPE has to be used, the only risk being that of missing
some small-scale features.  
   
\textit{Acknowledgements. }
Interactions with G. Spizzo, S. Cappello, and D.F. Escande are acknowledged. 
L. Salasnich, G. Serianni and Prof. J. Feder provided useful references. The referees made unvaluable
comments in order to improve the manuscript.   
This work was supported by the European Communities under the Contract of Association
between Euratom/ENEA.

\end{document}